\documentclass[fleqn,twoside]{article}
\usepackage{espcrc2,psfig}

\newlength{\figwidth}
\title{One-loop renormalization of heavy-light currents%
       \thanks{Talk presented by J. Harada.}}

\author{
  J. Harada\address{Dept. of Physics, Hiroshima University, 
                    Higashi-Hiroshima 739-8526, Japan} 
 \thanks{e-mail address:harada@theo.phys.sci.hiroshima-u.ac.jp },
  S. Hashimoto\address{High Energy Accelerator Reserch Organization (KEK), 
                       Tsukuba, Ibaraki 305-0801, Japan}, 
  K.-I. Ishikawa\address{Center for Computational Physics, 
                        University of Tsukuba, 
                        Tsukuba, Ibaraki 305-8577, Japan},
  A. S. Kronfeld\address{Theoretical Physics Department, 
                         Fermi National Accelerator Laboratory,
                         Batavia, Illinois 60510, USA},
  T. Onogi\address{Yukawa Institute for Theoritical Physics, 
                   Kyoto University, Kyoto 606-8502, Japan},
  N. Yamada$^{\rm b}$
}
\begin{document}

\begin{abstract}
 We calculate the mass dependent renormalization factors of heavy-light
 bilinears at one-loop order of perturbation theory, when the
 heavy quark is treated with the Fermilab formalism.
 We present numerical results for the Wilson and Sheikholeslami-Wohlert
 actions, with and without tree-level rotation.
 We find that in both cases our results smoothly interpolate from
 the static limit to the massless limit. 
 We also calculate the mass dependent Brodsky-Lepage-Mackenzie scale $q^*$, 
 with and without tadpole-improvement.
\end{abstract}

\maketitle

\section{INTRODUCTION}
 Although lattice QCD offers a nonperturbative method of calculating 
 weak matrix elements from first principles, in practice 
 a perturbative renormalization is also required to extract 
 the continuum quantities for heavy-light systems. 
 In this talk we discuss the renormalization of heavy-light vector and axial 
 vector currents.
 These currents are needed for heavy quark phenomenology, such as 
 the calculation of the decay constants and semi-leptonic form factors 
 of heavy-light mesons. 
 Here we calculate explicitly the mass dependent renormalization factors 
 of heavy-light currents at one-loop order, when the heavy quark is
 treated with the Fermilab formalism~\cite{EKM97}. 
 Results for the Wilson action have been
 obtained first in Ref.~\cite{K98} and preliminary results for clover
 action have been reported in previous lattice conferences~\cite{IAHMOY98}. 
 For tree-level improvement at order $1/m_Q$, we include so-called
 rotation term here.
 Tadpole-improved renormalization factors are also presented.
 We also calculate mass dependent Brodsky-Lepage-Mackenzie scale
 $q^*$~\cite{BLM83:LM93}, with and without tadpole-improvement.
 More details of this work will be given in Ref.~\cite{HHIKOY01}.

\section{ONE-LOOP RESULTS}
 The renormalization factors $Z_{J_\Gamma}$ of heavy-light currents
 are simply the ratio of the lattice and continuum radiative corrections:
 \begin{eqnarray}
  \indent
  Z_{J_\Gamma} = \frac{[Z_{2h}^{1/2}\Lambda_\Gamma Z_{2l}^{1/2}]^{cont}}
                     {[Z_{2h}^{1/2}\Lambda_\Gamma Z_{2l}^{1/2}]^{lat}},
 \end{eqnarray}
 where $Z_{2h}$ and $Z_{2l}$ are wave-function renormalization factors
 of the heavy and light quarks, and the vertex function $\Lambda_\Gamma$
 is the sum of one-particle irreducible three-point diagrams.
 We calculate explicitly $Z_A$ and $Z_V$ at one-loop order of
 perturbation theory. 

\begin{figure}[t]
    \begin{tabular}{c}
      \raisebox{0em}{\psfig{file=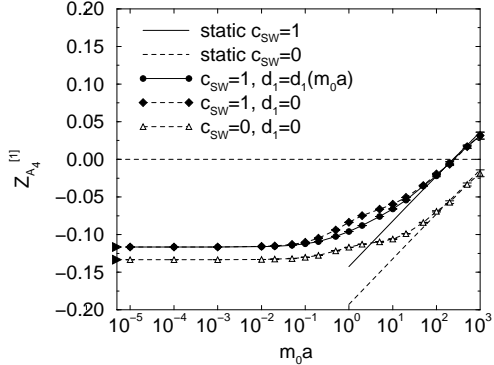,width=6.0cm}} 
  \end{tabular}
  \vspace{-1cm}
  \caption{One-loop renormalization coefficient $Z_{A_4}^{[1]}$ as a function
           of $am_0$.}
  \vspace{-0.7cm}
  \label{fig:ZA4}
\end{figure}

\begin{figure}[t]
    \begin{tabular}{c}
      \raisebox{0em}{\psfig{file=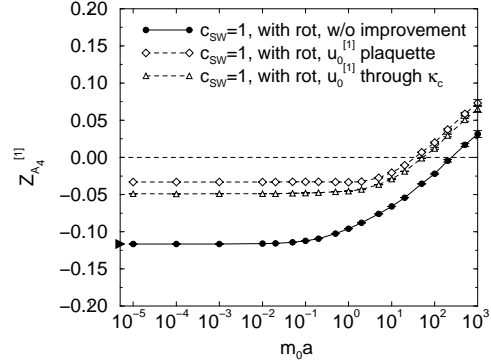,width=6.0cm}} 
  \end{tabular}
  \vspace{-1cm}
  \caption{Tadpole-improved one-loop renormalization coefficient 
           $Z_{A_4}^{[1]}$ as a function of $am_0$.}
  \vspace{-0.7cm}
  \label{fig:tadZA4}
\end{figure}

 In view of the mass dependence, we write 
 \begin{eqnarray}
  \indent
  e^{-m_1^{[0]}a/2} Z_{J_\Gamma} = 1 + \sum_{l=1}^\infty g_0^{2l} 
                                 Z_{J_\Gamma}^{[l]},
 \end{eqnarray}
 so that the $Z_{J_\Gamma}^{[l]}$ are only mildly mass dependent.
 Fig.~\ref{fig:ZA4} plots the full mass dependence of the renormalization
 factors for the axial vector current $Z_{A_4}^{[1]}$. 
 These numerical results are for 
 the SW action with and without rotation, and also for Wilson action
 without rotation. Our results agree with those previously obtained, for
 $c_{SW}=0$~\cite{K98} and for $c_{SW}=1,
 d_1=0$~\cite{IAHMOY98}~\footnote{The coefficient $d_1$ is field rotation
 parameter. See Ref.~\cite{EKM97}.}.  
 We find that in both cases our results smoothly
 interpolate from the static to massless limit.
 The resulting analytical expressions are in Ref.~\cite{HHIKOY01}.
 Fig.~\ref{fig:tadZA4} plots the tadpole-improved renormalization factor
 for $Z_{A_4}^{[1]}$. From this figure, we can see that 
 tadpole-improvement significantly reduces the
 one-loop coefficients of renormalization factors.
 Results for $Z_{A_i}$ and $Z_{V_{4,i}}$ are given in Ref.~\cite{HHIKOY01}.

 The slope of our mass-dependent renormalization factors in the massless
 limit is related to the improvement 
 coefficients $b_J$ and $c_J$~\cite{LSSW96}.
 We find 
 \begin{eqnarray}
  b_V^{[1]} &=& 0.153239(14), \\
  b_A^{[1]} &=& 0.152189(14), \\
  c_V^{[1]} &=& -0.016332(7), \\
 \indent 
  c_A^{[1]} &=& -0.0075741(15).
 \end{eqnarray}
 These results agree perfectly with Ref.~\cite{SW97}. 
 We also obtain by subtracting the integrands first, 
 \begin{eqnarray}
 \indent  b_V^{[1]} - b_A^{[1]} = 0.0010444(16)
 \end{eqnarray}
 which is more accurate than the difference of the two numbers quoted above.
 We find our one-loop result of $b_V-b_A$ are far from
 nonperturbative calculations~\cite{LSSW97:BGLS01}. 

\section{SETTING THE SCALE}
\begin{figure}[t]
    \begin{tabular}{c}
      \raisebox{0em}{\psfig{file=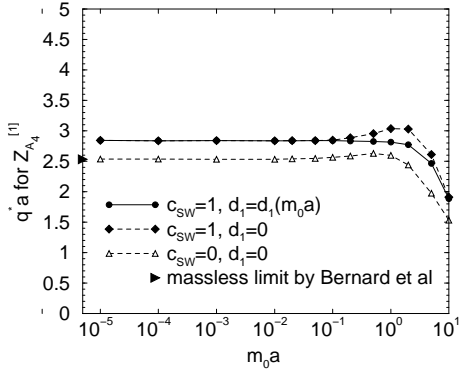,width=6.0cm}} 
  \end{tabular}
  \vspace{-1cm}
  \caption{Brodsky-Lepage-Mackenzie scale $q^*$ for $Z_{A_4}^{[1]}$ as a
 function of $am_0$.}
  \vspace{-0.5cm}
  \label{fig:qstarA4}
\end{figure}

\begin{figure}[t]
    \begin{tabular}{c}
      \raisebox{0em}{\psfig{file=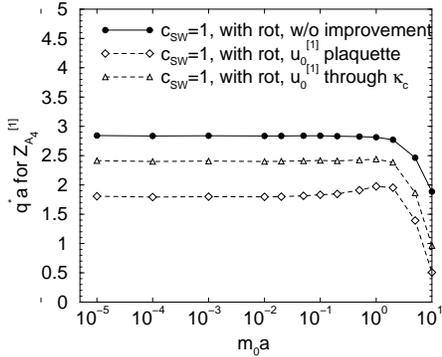,width=6.0cm}} 
  \end{tabular}
  \vspace{-1cm}
  \caption{Tadpole-improved Brodsky-Lepage-Mackenzie scale $q^*$ for 
           $Z_{A_4}^{[1]}$ as a function of $am_0$.}
  \vspace{-0.7cm}
  \label{fig:tadqstarA4}
\end{figure}

 The typical gluon momentum $q^*$ in the $V$-scheme, as suggested by
 Brodsky, Lepage and Mackenzie (BLM), is defined by~\cite{BLM83:LM93}
 \begin{equation}
  \indent
  \ln ({q^*}^2) \equiv \frac{\int d^4q f(q) \ln (q^2)}{\int d^4 qf(q)},
   \label{eq:qstar}
 \end{equation}
 where $q$ is the momentum of gluon, and 
 the form $\int d^4q f(q)$ is the one-loop integral for a
 particular renormalization constant, for example, 
 $\int d^4q f(q) = Z_{J_\Gamma}^{[1]}$.
 Previously $q^*$ has been calculated for the light-light current~\cite{BGM99}
 \cite{CLV98} and the static-light current~\cite{HH94}.
 Here we calculate the mass dependent $q^*$ for the heavy-light current.
 Results are plotted in Fig.~\ref{fig:qstarA4}. 
 For Wilson action case, our
 results agree with Ref.~\cite{BGM99} in the massless limit. 
 From Fig.~\ref{fig:qstarA4}, 
 we can see that the mass dependence of $q^*$ is weak from
 massless limit to $m_0a \sim 1$, 
 especially for clover with rotation case. 
 The original BLM prescription of $q^*$ breaks down at larger masses, 
 because $Z_{J_\Gamma}^{[1]}$(denominator in
 Eq.~(\ref{eq:qstar})) goes through zero at there.
 A prescription for $q^*$ in this case is given in Ref.~\cite{HLM01}.
 We also calculate tadpole-improved $q^*$ and results are plotted in
 Fig.~\ref{fig:tadqstarA4}. 
 We can see that plaquette tadpole-improvement significantly reduces $q^*$,
 on the other hand, the reduction is rather small for $\kappa_c$
 tadpole-improvement. We summarize the results in the massless limit in
 Table.~\ref{tab:one-loop}.

 We can also obtain the BLM scale for improvement coefficients $b_J$ and
 $c_J$~\cite{LSSW96}. Then it is interesting to compare BLM perturbation
 theory with non-perturbative calculations of these
 coefficients~\cite{LSSW97:BGLS01}\cite{C01}.
 We will present these results for $q^*$ and the mentioned comparison 
 in another publication~\cite{HHKO01no2}.


\section{CONCLUSIONS}
 We have obtained one-loop results of $Z_A$ and $Z_V$ with tree-level
 rotation, which should be useful for lattice calculations of $f_B$ and of
 form factors for $B\rightarrow \pi l \nu$.
 We have also obtained the BLM scale $q^*$ for arbitrary masses, which
 should reduce the uncertainty of one-loop calculations.

\begin{table}[t]
\caption{One-loop $Z$-factor and BLM scale $q^*$ in the massless limit
 for clover (upper
 row) or Wilson (lower row) action with tadpole-improvement.}
\begin{center}
\small
\begin{tabular}{cccc}
\hline\hline
 & \hspace{-0.5cm}no improvement & plaquette & through $\kappa_c$ \\
\hline
$Z_{A}^{[1]}$ & $-0.116457(2)$ & $-0.033124(2)$ & $-0.048938(2)$ \\
              & $-0.133375(2)$ & $-0.050042(2)$ & $-0.024803(4)$ \\
$Z_{V}^{[1]}$ & $-0.129430(2)$ & $-0.046097(2)$ & $-0.061911(2)$ \\
              & $-0.174086(2)$ & $-0.090752(2)$ & $-0.065514(4)$ \\
$q_{Z_A}^*a$  & $ 2.839$       &  $1.802$       &  $2.408$       \\
              & $ 2.533$       &  $1.550$       &  $2.316$       \\
$q_{Z_V}^*a$  & $ 2.845$       &  $2.060$       &  $2.503$       \\
              & $ 2.370$       &  $1.700$       &  $2.052$       \\
\hline\hline
\end{tabular}
\end{center}
\label{tab:one-loop}
\vspace{-0.8cm}
\end{table}

\section*{Acknowledgments}
S.H. and T.O. are supported by the Grant-in-Aid of the Japanese Ministry 
of Education, (Nos.11740162, No.12640279).
K.-I.I. and N.Y. are supported by the JSPS Research Fellowships. 
Fermilab is operated by Universities Research Association Inc., under
contract with the U.S. Department of Energy.



\end{document}